\documentclass[twoside]{dis08}
\usepackage[latin1]{inputenc}
\usepackage{graphicx,epsfig,color}
\usepackage{wrapfig}
\usepackage{amssymb,amsmath,array}

\begin{document}

\title{Potential PDF sensitivity at LHCb}

\author{Ronan McNulty on behalf of the LHCb Collaboration
%
\thanks{The author would like to acknowledge the support of Science Foundation Ireland}
%
\vspace{.3cm}\\
%
School of Physics, 
University College Dublin, \\
Belfield, Dublin 4. Ireland.
%
}

\maketitle

\begin{abstract}
A review of the potential sensitivity of the LHCb
experiment to the parton distribution functions is given.
Studies of dimuon events coming from Z, W and low mass Drell Yan production
are presented and compared to MSTW theoretical predictions.

\end{abstract}

\section{Introduction}

LHCb~\cite{lhcb}, one of the four main experiments at the LHC, is designed to make measurements
of CP violation and rare B decays in the forward region
($1.9 \leq \eta \leq 4.9$). However, it can also make precision electroweak
measurements at high rapidities which enable it to
probe the proton parton distribution functions (PDFs) in hitherto 
unmeasured regions. 

Part of LHCb's pseudorapidity range ($1.9 \leq \eta \leq 2.5$) overlaps
that of ATLAS and CMS, however measurements made at $\eta > 2.5$ are
unique to LHCb.  Other distinguishing features
are the lower luminosity which reduces the number of pile-up events
(one year of nominal running should collect 2 fb$^{-1}$ of data), 
and the ability to trigger on low transverse momentum muons, $p_T$, down to
1 GeV/c.

We describe results of simulation studies~\cite{lhcbcomputing}
that show LHCb will be able
to trigger on and reconstruct Z and W bosons with high efficiency and purity.
Phenomenologically LHCb differs from the other experiments
since the production of W and Z here occurs
predominantly through sea-valence interactions, in contrast to the sea-sea
interactions in the central region~\cite{wzlhc}.
We also present studies of the Drell-Yan production of low-mass muon pairs
which suggest that LHCb can isolate such events down to invariant masses,
$Q^2$, of 5 GeV/c$^2$.  

We discuss the implications that such measurements will have on
constraining the proton PDFs: 
in particular, the Drell-Yan final states allow us to probe values of the
parton momentum fraction, $x$, down to
$2\times 10^{-6}$, a previously unexplored region.

\section{Measurements of Z production}
\label{zprod}

Z boson decays to dimuons possess a distinctive signature
that allow
them to be triggered~\cite{lhcbtrigger} and reconstructed with high efficiency and purity.
The dimuon trigger at LHCb is very loose and requires two muon candidates whose summed
transverse momentum exceeds 1.6 GeV/c.  Offline selections suppress background
processes by requiring: that both muons candidates have little activity in the
hadronic calorimeter; that the higher and lower transverse momentum muons are
greater than 20 GeV/c and 15 GeV/c respectively; that both have impact parameters consistent
with the primary interaction; and their combined invariant mass lies within 20 GeV/c$^2$
of the Z mass.  
Backgrounds have been studied~\cite{pythia} 
from other electroweak processes 
($Z \rightarrow \tau \tau$ decays, 
$Z \rightarrow bb$ decays, W + jet production, WW production); quark 
production ($bb$, $cc$, $tt$, and single top); and pions and kaons misidentified as muons.

For Z bosons decaying to two muons which are within the LHCb acceptance,
the selection efficiency is $91\pm 1$\%
and the purity is estimated to be $97\pm 3$ \% with
the dominant background ($3\pm3$\%) being mis-identified hadrons whose contribution
will be precisely determined from data.
After just 50 pb$^{-1}$ of data, the cross-section measurement will be limited by
systematic uncertainties to about 1\%, assuming 
the machine luminosity can be precisely determined.
The dominant uncertainty in the theoretical prediction comes from knowledge of
the PDFs and is from 2-5\% depending on rapidity, and
thus an experimental measurement of the Z cross-section will constrain the PDFs.

\begin{wrapfigure}{r}{0.65\columnwidth}
\includegraphics[width=9cm]{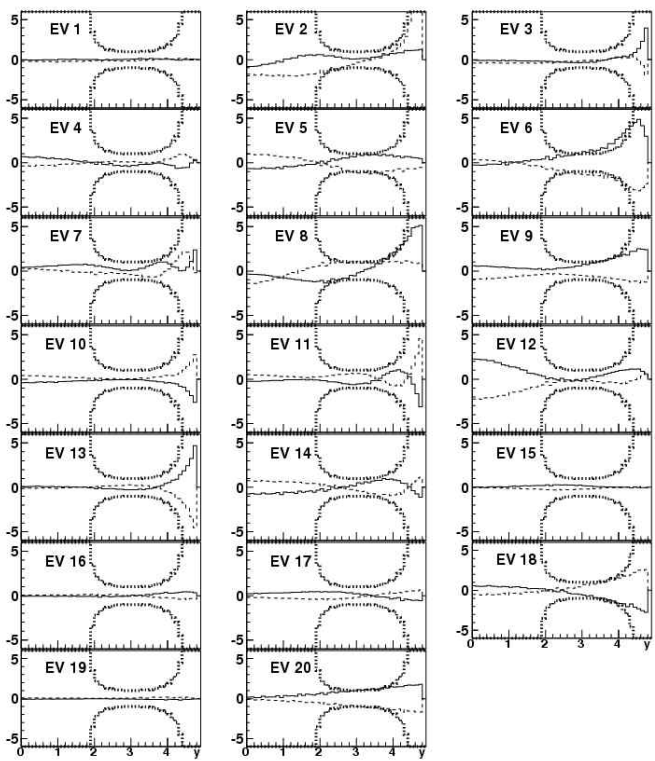}
\caption{A comparison of the theoretical and experimental precision
on the differential cross-section for Z boson production as a function
of rapidity, $y$, using
1fb$^{-1}$ of LHCb data for each of the
20 eigenvectors in the MSTW07 parametrisation.
The $+1 (-1) \sigma$ uncertainties are indicated with a
continuous (dashed) line.  The experimental uncertainty is indicated
with a dotted line.}
\label{zpred}
\end{wrapfigure}

The sensitivity of the PDFs to a measurement of the Z cross-section can be seen
in Figure~\ref{zpred} where we have plotted the variation in the 
differential cross-section produced using the MCFM generator~\cite{mcfm}
with
$\pm 1\sigma$ variations in
each of the 20 eigenvectors~\cite{ev} in the MSTW07 parametrisation~\cite{mstw07}.  
Superimposed upon these
are the expected experimental uncertainties with 1fb$^{-1}$ of data, assuming 
that the luminosity is known.

Current estimates for the precision on a luminosity determination are from 1 to 3\%
using the beam-gas technique~\cite{beamgas}, or by measuring the elastic two-photon 
production of leptons~\cite{twophoton}.  
However, even in the absence of a precise luminosity determination, information about 
the PDFs can be derived from
the {\it shape} of the differential distribution $d\sigma\over dy$; 
only the 
overall normalisation is fixed by the luminosity.

\section{W production}
\label{wprod}
W boson decays to a lepton plus neutrino have previously been observed at
hadron colliders by observing high momentum leptons in association with
missing transverse energy, a quantity that can not be measured well at
LHCb since it is only instrumented in the forward region.  
Nonetheless, our studies indicate
that the presence of a single high transverse momentum muon which makes up
the dominant observed activity in the event, is sufficent to select
a rather clean samples of Ws.  

First we require the presence of a high transverse momentum muon, \mbox{$p_T>30{\rm GeV/c}$}.
Defining the asymmetry, $A$, between two quantities $x$ and $y$ to be 
\mbox{$A(x,y)={x-y\over x+y}$} we compare the muon momentum, $p^\mu$, to the summed
momentum of every other charged particle detected in LHCb, $p^{\rm rest}$, and
require that $A(p^\mu,p^{\rm rest})>0.85$.  Whereas the signal is 
peaked towards 1, backgrounds coming from semi-leptonic b decays or mis-identified
pions are strongly peaked towards -1.  The combination of these requirements has
an overall selection efficiency of about 33\% and a purity of 83\%.  
Again, the dominant 
background is due to misidentifying a pion as a muon.
It is noteworthy that in LHCb, Z decays form a small (3\%) but irreducible background, 
caused when only one of the dimuons enters the LHCb
acceptance.

A full study of systematic errors is ongoing though it is expected that a similar
experimental precision to the Z channel (1\%) can be obtained.
The comparison to the theoretical prediction is complicated by the fact
that the W boson itself can not be reconstructed.  Consequently we have compared
our distributions to the theoretical expectations for the differential distributions
expressed as a function of lepton rapidities.  Once again knowledge of 
the PDFs are the dominant uncertainties (ranging from 2\% to 4\% for MSTW07) and so
the experimental data can constrain the PDFs.

The uncertainty due to luminosity can be removed by comparing ratios 
and asymmetries of 
the $W^+,W^-$ and $Z$ differential cross-sections in such a way as
to minimise or maximise the dependence on the PDFs~\cite{ratios}.
The ratio $R_{ZW}={\sigma_Z\over\sigma_{W+}+\sigma_{W-}}$
is then almost independent of PDF uncertainties and thus constitutes a 
precise electroweak prediction for the LHC.  
The ratio $R_{\pm}={\sigma_{W+}\over\sigma_{W-}}$ has an uncertainty that rises
from $<1\%$ at $y=0$ to
20\% at $y=4.5$ 
and at high rapidities, which only LHCb can measure, 
is sensitive to the large-$x$ $u/d$ ratio of partons.
Measuring the asymmetry $A(W+,W-)$ will greatly constrain the PDFs since
the current theoretical uncertainty on this quantity is about 10\% 
at all rapidities.

\section{Low invariant mass Drell-Yan production}
\label{dy}

Dimuons produced below the Z resonance can be recorded down to an invariant mass
of \mbox{1.6 Gev/c$^2$} using the dimuon trigger.  However, isolating the signal from
the order-of-magnitude larger backgrounds coming from semileptonic heavy quark
decays and pion misidentification is challenging.
Considering the topology of the signal leads us to consider five observables:
the momenta of each of the muons ($p^{\mu 1},p^{\mu 2}$); the summed momenta
of all particles found within a cone of one unit of phi and pseudorapidity
about each muon ($p^{cone1},p^{cone2}$); and the summed momenta of any other
activity in the event ($p^{rest}$).  
Five asymmetries are defined:
\mbox{$A(p^{\mu 1},p^{\mu 2})$};
\mbox{$A(p^{\mu 1},p^{cone1})$};
\mbox{$A(p^{\mu 2},p^{cone2})$};
\mbox{$A(p^{\mu 1}+p^{\mu 2},p^{rest})$};
\mbox{$A(p^{cone1}+p^{cone2},p^{rest})$};
whose shapes are almost independent of the invariant mass of the muons.
These variables are combined to form a likelihood.  Cutting progressively
harder on this likelihood increases the purity of the signal though at the expense
of efficiency.  
Our studies show that above a dimuon invariant mass of 10 GeV/c$^2$, purities
above 95\% can be obtained retaining between 15\% and 50\% of triggered
signal events, so that the total number of selected events per mass bin
is roughly constant.
Below 10 GeV/c$^2$ pure signals can not be obtained although measurements of
the cross-section may be possible down to 5 GeV/c$^2$ where 5\% of the signal 
events can be retained with a purity of 70\%.

\begin{wrapfigure}{r}{0.39\columnwidth}
\includegraphics[height=6.5cm, bb=0 27 568 450]{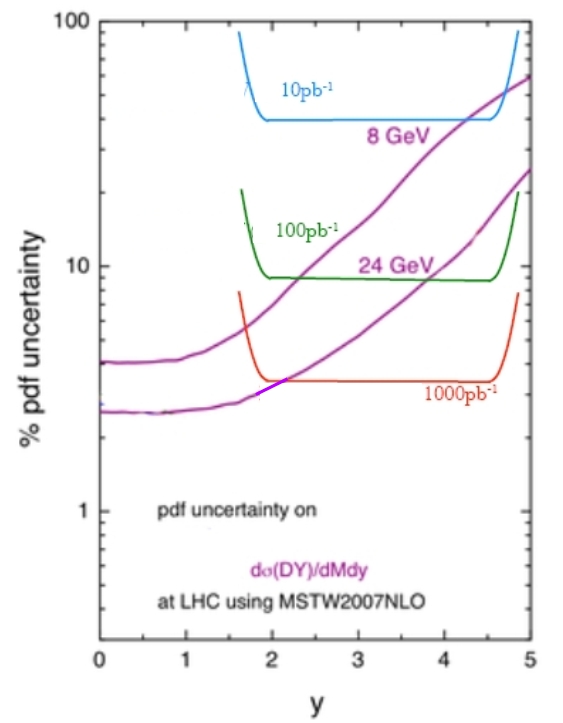}
\caption{
Two theoretical curves from \cite{plenary} showing the PDF percentage uncertainty
as a function of rapidity for $Q^2$ of 8 and 24 GeV/c$^2$.
Superimposed are the experimental statistical precision 
(which is almost independent of $Q^2$)
that can be
obtained with 10, 100 and \mbox{1000 pb$^{-1}$} of data.
}
\label{dyth}
\end{wrapfigure}


The experimental precision obtainable on the
double differential distribution, $d^2\sigma/dQ^2dy$,
is compared with
the current theoretical uncertainties due to
the PDFs in Figure~\ref{dyth}.
With 1 fb$^{-1}$ of data, 
and with bins of 0.1 in rapidity and 2 GeV/c$^2$ in mass,
a statistical precision of a few percent per bin is obtainable,
though systematic uncertainties remain to be determined and 
could be larger at
low masses where the backgrounds are greater.
The theoretical uncertainties at low invariant masses is substantial
and driven by knowledge of the small-$x$ gluon, approaching
60\% at high rapidities for $Q^2=8$GeV/c$^2$.  

Within the first year of data-taking, LHCb should thus start to
constrain
the gluon PDF and make significant improvements on current
knowledge after a few years.  At high rapidities and values of $Q^2=5$GeV/c$^2$,
LHCb will be able to probe $x$ values down to $2\times 10^{-6}$ where the
PDFs are currently unknown.

\section{Conclusions}
\label{conc}

The LHCb experiment can access a unique range of $x$ and Q$^2$ at the LHC. 
Ratios and asymmetries of the 
W and Z cross-sections will be experimentally determined to 1\%
constraining the PDFs which currently have a theoretical uncertainty
of 2-5\% at $Q^2\approx 90$GeV/c$^2$.
Studies of low invariant mass Drell-Yan dimuon states indicate that LHCb can
rapidly improve the knowledge of PDFs down to 
$x$ values of $2\times 10^{-6}$ and Q$^2$ of 5 GeV/c$^2$ 
and 
significantly improve our knowledge of the small-$x$ gluon.


 

\begin{footnotesize}



%

\end{footnotesize}


\end{document}